# Graphyne and Borophene as Nanoscopic Materials for Electronics

C. M. Krowne
Electromagnetics Technology Branch, Electronics Science & Technology Division,
Naval Research Laboratory, Washington, DC 20375

**Abstract**

Discussions based upon rigorous derivations show the validity range of the analogy between solid state materials like graphene which possess K symmetry crystallographic points in **k**-space, and the relativistic solutions for massive and low mass particles associated with the Dirac equation. Both eigenenergies and eigenvectors are examined for the nonrelativistic solutions of the Schrodinger equation using the tight-binding method, and the relativistic solutions of the Dirac equation. Implications for exploring new materials are drawn from the results. It is concluded Dirac materials are unlikely to fulfill the needs of transistor action materials, but two prime candidates which may satisfy those needs for 2D future electronics are proposed, graphyne and borophene.

**Introduction**

It is worthwhile to assess what can be understood from the present day association of graphene with relativistic attributes. Our interest here originally arose out of studying the metal-insulator transition properties of graphene [1] and a nm scale ruthenium system [2]. Most widely known is the labeling of its particular K-space graphene symmetry points with the terminology Dirac. Electrons carrying the current for a graphene monoatomic sheet have very high mobility, and under an applied electric field, attain a very high velocity compared to more conventional semiconductors, whether monoatomic like Si and Ge, or compounds like GaAs and GaN, and many present day variants having three or more atomic constituents. The analogy between graphene like materials and the Dirac equation carries over to basically massless Dirac Fermions in the relativistic sense. This is because the transport in graphene, based on π-electrons from the carbon atoms, are very low mass compared to the atoms themselves. These issues are examined in the following treatment. First is addressed a formulation for electronic bandstructure which yields analytical results for 2D hexagonal materials. Second, from this are examined the eigenenergies, the Fermi velocities, and the relevant overlap and hopping integrals. Next attention is turned toward the graphene and relativistic QED type particles eigenvectors, namely their spinors in 2-spinor (third section) and 4-spinor forms (fourth section). Lastly, conclusions are drawn, and what could be the implications for other 2D materials in the near future. It is discovered that most likely non-Dirac properties for 2D materials may be most desirable for many transport electronic devices if switching action is desirable, leading to the suggestion that materials development occur for graphyne and borophene.

**Formulation of the Bandstructure Equations in a Tractable Tight-Binding Format**

Because there are two atom types in 2D hexagonal materials, corresponding to the two atom types within the unit cell Bravais lattice, the total electronic wavefunction must be the superposition of wavefunctions $\psi_\mathbf{k}^A(\mathbf{r})$ and $\psi_\mathbf{k}^B(\mathbf{r})$ [3]. That is,

$$\psi_\mathbf{k}(\mathbf{r}) = \sum_{\ell_{sub}} a_\mathbf{k}^{l_{sub}} \psi_\mathbf{k}^{l_{sub}}(\mathbf{r}) = a_\mathbf{k} \psi_\mathbf{k}^A(\mathbf{r}) + b_\mathbf{k} \psi_\mathbf{k}^B(\mathbf{r}) \qquad (1)$$

Index $l_{sub}$ is the sublattice atom index type, and can equal generally any number of atom types contained within the Bravais unit cell lattice.

Only way to convert the spatially varying Schrodinger equation,

$$H(\mathbf{r})\psi_\mathbf{k}(\mathbf{r}) = \varepsilon_\mathbf{k} \psi_\mathbf{k}(\mathbf{r}) \qquad (2)$$

with Hamiltonian $H(\mathbf{r})$,

$$H(\mathbf{r}) = -\frac{\hbar^2}{2m_e}\nabla^2 + V(\mathbf{r}) \qquad (3)$$

[$V(\mathbf{r})$ is the total scalar potential energy at a point $\mathbf{r}$ that the electron experiences in the periodic lattice consisting of the atoms, each atom made up of its electrons and its positive atomic core] into a form which can be numerically evaluated, is to eliminate the vector spatial variable $\mathbf{r}$ by integrating it out. By (1), total wavefunction consists as the sum over the sublattice atoms. For carbon, that is two types labeled A and B. Equation (1) in vector form appears as

$$\psi_\mathbf{k}(\mathbf{r}) = \begin{bmatrix} \psi_\mathbf{k}^A(\mathbf{r}) & \psi_\mathbf{k}^B(\mathbf{r}) \end{bmatrix} \begin{bmatrix} a_\mathbf{k} \\ b_\mathbf{k} \end{bmatrix} \qquad (4)$$

Its Hermitian conjugate is

$$\psi_\mathbf{k}^\dagger(\mathbf{r}) = [\psi_\mathbf{k}(\mathbf{r})]^\dagger = \left\{\begin{bmatrix} \psi_\mathbf{k}^A(\mathbf{r}) & \psi_\mathbf{k}^B(\mathbf{r}) \end{bmatrix} \begin{bmatrix} a_\mathbf{k} \\ b_\mathbf{k} \end{bmatrix}\right\}^\dagger = \begin{bmatrix} a_\mathbf{k} \\ b_\mathbf{k} \end{bmatrix}^\dagger \begin{bmatrix} \psi_\mathbf{k}^A(\mathbf{r}) & \psi_\mathbf{k}^B(\mathbf{r}) \end{bmatrix}^\dagger \qquad (5a)$$

or

$$\psi_\mathbf{k}^\dagger(\mathbf{r}) = [\psi_\mathbf{k}(\mathbf{r})]^\dagger = \begin{bmatrix} (a_\mathbf{k})^* & (b_\mathbf{k})^* \end{bmatrix} \begin{bmatrix} \{\psi_\mathbf{k}^A(\mathbf{r})\}^* \\ \{\psi_\mathbf{k}^B(\mathbf{r})\}^* \end{bmatrix} = \begin{bmatrix} a_\mathbf{k}^* & b_\mathbf{k}^* \end{bmatrix} \begin{bmatrix} \psi_\mathbf{k}^{A\,*}(\mathbf{r}) \\ \psi_\mathbf{k}^{B\,*}(\mathbf{r}) \end{bmatrix} \qquad (5b)$$

Multiplying the Schrodinger equation from the left by $\psi_\mathbf{k}^\dagger(\mathbf{r})$, (2) becomes

$$\psi_\mathbf{k}^\dagger(\mathbf{r}) H(\mathbf{r}) \psi_\mathbf{k}(\mathbf{r}) = \varepsilon_\mathbf{k} \psi_\mathbf{k}^\dagger(\mathbf{r}) \psi_\mathbf{k}(\mathbf{r}) \qquad (6)$$

Integrating over two dimensional space for our 2D crystalline system,

$$\iint d^2\mathbf{r} \, \psi_\mathbf{k}^\dagger(\mathbf{r}) H(\mathbf{r}) \psi_\mathbf{k}(\mathbf{r}) = \varepsilon_\mathbf{k} \iint d^2\mathbf{r} \, \psi_\mathbf{k}^\dagger(\mathbf{r}) \psi_\mathbf{k}(\mathbf{r}) \qquad (7)$$

Substitute in $\psi_\mathbf{k}^\dagger(\mathbf{r})$ and $\psi_\mathbf{k}(\mathbf{r})$ from (4) and (5b) into (7),

$$\iint d^2\mathbf{r} \begin{bmatrix} a_\mathbf{k}^* & b_\mathbf{k}^* \end{bmatrix} \begin{bmatrix} \psi_\mathbf{k}^{A\,*}(\mathbf{r}) \\ \psi_\mathbf{k}^{B\,*}(\mathbf{r}) \end{bmatrix} H \begin{bmatrix} \psi_\mathbf{k}^A(\mathbf{r}) & \psi_\mathbf{k}^B(\mathbf{r}) \end{bmatrix} \begin{bmatrix} a_\mathbf{k} \\ b_\mathbf{k} \end{bmatrix} = \varepsilon_\mathbf{k} \iint d^2\mathbf{r} \begin{bmatrix} a_\mathbf{k}^* & b_\mathbf{k}^* \end{bmatrix} \begin{bmatrix} \psi_\mathbf{k}^{A\,*}(\mathbf{r}) \\ \psi_\mathbf{k}^{B\,*}(\mathbf{r}) \end{bmatrix} \begin{bmatrix} \psi_\mathbf{k}^A(\mathbf{r}) & \psi_\mathbf{k}^B(\mathbf{r}) \end{bmatrix} \begin{bmatrix} a_\mathbf{k} \\ b_\mathbf{k} \end{bmatrix} \qquad (8)$$

Knowing that quantum mechanically the Hamiltonian $H$ operator is scalar here, performing outer products, allows identification of two matrix operators,

$$\mathcal{H}_k(\mathbf{r}) = \begin{bmatrix} \psi_k^{A*}(\mathbf{r}) \\ \psi_k^{B*}(\mathbf{r}) \end{bmatrix} H \begin{bmatrix} \psi_k^A(\mathbf{r}) & \psi_k^B(\mathbf{r}) \end{bmatrix} = \begin{bmatrix} \psi_k^{A*}(\mathbf{r})H(\mathbf{r})\psi_k^A(\mathbf{r}) & \psi_k^{A*}(\mathbf{r})H(\mathbf{r})\psi_k^B(\mathbf{r}) \\ \psi_k^{B*}(\mathbf{r})H(\mathbf{r})\psi_k^A(\mathbf{r}) & \psi_k^{B*}(\mathbf{r})H(\mathbf{r})\psi_k^B(\mathbf{r}) \end{bmatrix} \quad (9a)$$

$$S_k(\mathbf{r}) = \begin{bmatrix} \psi_k^{A*}(\mathbf{r}) \\ \psi_k^{B*}(\mathbf{r}) \end{bmatrix} \begin{bmatrix} \psi_k^A(\mathbf{r}) & \psi_k^B(\mathbf{r}) \end{bmatrix} = \begin{bmatrix} \psi_k^{A*}(\mathbf{r})\psi_k^A(\mathbf{r}) & \psi_k^{A*}(\mathbf{r})\psi_k^B(\mathbf{r}) \\ \psi_k^{B*}(\mathbf{r})\psi_k^A(\mathbf{r}) & \psi_k^{B*}(\mathbf{r})\psi_k^B(\mathbf{r}) \end{bmatrix} \quad (9b)$$

Equations (9) represent the Hamiltonian tested by the A and B atom total wavefunctions $\mathcal{H}_k(\mathbf{r})$, and the same and mixed products of the total A and B atom wavefunctions $S_k(\mathbf{r})$ (self matrix of the sublattice wavefunctions). Notice that the second equation of (9) can be obtained from the first by formally letting $H \to 1$. Now integral equation (8) can be written compactly as

$$\iint d^2\mathbf{r} \begin{bmatrix} a_k^* & b_k^* \end{bmatrix} \mathcal{H}_k(\mathbf{r}) \begin{bmatrix} a_k \\ b_k \end{bmatrix} = \varepsilon_k \iint d^2\mathbf{r} \begin{bmatrix} a_k^* & b_k^* \end{bmatrix} S_k(\mathbf{r}) \begin{bmatrix} a_k \\ b_k \end{bmatrix} \quad (10)$$

Matrix operations and integration order can be switched in (10), so the double integral can be pulled past the coefficient row or column vectors.

$$\begin{bmatrix} a_k^* & b_k^* \end{bmatrix} \left\{ \iint d^2\mathbf{r}\, \mathcal{H}_k(\mathbf{r}) \right\} \begin{bmatrix} a_k \\ b_k \end{bmatrix} = \varepsilon_k \begin{bmatrix} a_k^* & b_k^* \end{bmatrix} \left\{ \iint d^2\mathbf{r}\, S_k(\mathbf{r}) \right\} \begin{bmatrix} a_k \\ b_k \end{bmatrix} \quad (11)$$

The integrated out Hamiltonian and self matrices, are identified as

$$\bar{\mathcal{H}}_k = \iint d^2\mathbf{r}\, \mathcal{H}_k(\mathbf{r}) \quad ; \quad \bar{S}_k = \iint d^2\mathbf{r}\, S_k(\mathbf{r}) \quad (12)$$

Inserting (9) into (12) gives

$$\bar{\mathcal{H}}_k = \iint d^2\mathbf{r}\, \mathcal{H}_k(\mathbf{r}) = \iint d^2\mathbf{r} \begin{bmatrix} \psi_k^{A*}(\mathbf{r}) \\ \psi_k^{B*}(\mathbf{r}) \end{bmatrix} H \begin{bmatrix} \psi_k^A(\mathbf{r}) & \psi_k^B(\mathbf{r}) \end{bmatrix}$$

$$= \begin{bmatrix} \iint d^2\mathbf{r}\, \psi_k^{A*}(\mathbf{r})H(\mathbf{r})\psi_k^A(\mathbf{r}) & \iint d^2\mathbf{r}\, \psi_k^{A*}(\mathbf{r})H(\mathbf{r})\psi_k^B(\mathbf{r}) \\ \iint d^2\mathbf{r}\, \psi_k^{B*}(\mathbf{r})H(\mathbf{r})\psi_k^A(\mathbf{r}) & \iint d^2\mathbf{r}\, \psi_k^{B*}(\mathbf{r})H(\mathbf{r})\psi_k^B(\mathbf{r}) \end{bmatrix} \quad (13a)$$

$$\bar{S}_k = \iint d^2\mathbf{r}\, S_k(\mathbf{r}) = \iint d^2\mathbf{r} \begin{bmatrix} \psi_k^{A*}(\mathbf{r}) \\ \psi_k^{B*}(\mathbf{r}) \end{bmatrix} \begin{bmatrix} \psi_k^A(\mathbf{r}) & \psi_k^B(\mathbf{r}) \end{bmatrix}$$

$$= \begin{bmatrix} \iint d^2\mathbf{r}\, \psi_k^{A*}(\mathbf{r})\psi_k^A(\mathbf{r}) & \iint d^2\mathbf{r}\, \psi_k^{A*}(\mathbf{r})\psi_k^B(\mathbf{r}) \\ \iint d^2\mathbf{r}\, \psi_k^{B*}(\mathbf{r})\psi_k^A(\mathbf{r}) & \iint d^2\mathbf{r}\, \psi_k^{B*}(\mathbf{r})\psi_k^B(\mathbf{r}) \end{bmatrix} \quad (13b)$$

so that (11) can be rewritten as

$$\begin{bmatrix} a_k^* & b_k^* \end{bmatrix} \bar{\mathcal{H}}_k \begin{bmatrix} a_k \\ b_k \end{bmatrix} = \varepsilon_k \begin{bmatrix} a_k^* & b_k^* \end{bmatrix} \bar{S}_k \begin{bmatrix} a_k \\ b_k \end{bmatrix} \quad (14)$$

Collecting terms on the left-hand-side,

$$\begin{bmatrix} a_{\mathbf{k}}^* & b_{\mathbf{k}}^* \end{bmatrix} \{\bar{\mathcal{H}}_{\mathbf{k}} - \varepsilon_{\mathbf{k}} \bar{S}_{\mathbf{k}}\} \begin{bmatrix} a_{\mathbf{k}} \\ b_{\mathbf{k}} \end{bmatrix} = 0 \qquad (15)$$

Stripping off the left row matrix hitting the remaining double product,

$$\{\bar{\mathcal{H}}_{\mathbf{k}} - \varepsilon_{\mathbf{k}} \bar{S}_{\mathbf{k}}\} \begin{bmatrix} a_{\mathbf{k}} \\ b_{\mathbf{k}} \end{bmatrix} = 0 \qquad (16)$$

what is present is recognizable from linear matrix theory, as our implicit determinantal equation. That is, in order for (16) to have a solution for any arbitrary column matrix $\mathbf{S}_{sub}$ of the sublattice coefficients weighting the two types of atoms available, A and B,

$$\mathbf{S}_{sub} = \begin{bmatrix} a_{\mathbf{k}} \\ b_{\mathbf{k}} \end{bmatrix} \qquad (17)$$

with (16) streamlined to

$$\{\bar{\mathcal{H}}_{\mathbf{k}} - \varepsilon_{\mathbf{k}} \bar{S}_{\mathbf{k}}\} \mathbf{S}_{sub} = 0 \qquad (18)$$

the determinant of the matrix hitting $\mathbf{S}_{sub}$ must be zero, giving the secular equation

$$\det\{\bar{\mathcal{H}}_{\mathbf{k}} - \varepsilon_{\mathbf{k}} \bar{S}_{\mathbf{k}}\} = 0 \qquad (19)$$

This is the solution for a non-trivial $\mathbf{S}_{sub}$ when it is not null, when at least one of its elements are not zero. For a given $\mathbf{k}$ value, (19) will give two energy eigenvalue solutions $\varepsilon_{\mathbf{k}}^\lambda$, for $\lambda = 1, 2$ bands, associated with the fact that two sublattices of atoms exist, the A and B sublattices for graphene or some other two atom Bravais unit cell in real space. In general, the bands are labeled

$$\lambda = 1, 2, 3, \cdots, N_{ac} \qquad (20)$$

where $N_{ac}$ is the number of atoms inside the Bravais unit cell.

The problem may be generalized for any number of atoms in the Bravais unit cell. Clearly, secular equation (19) is already general. However, (1) must be generalized to

$$\psi_{\mathbf{k}}(\mathbf{r}) = \sum_{\ell_{sub}=1}^{N_{ac}} a_{\mathbf{k}}^{\ell_{sub}} \psi_{\mathbf{k}}^{\ell_{sub}}(\mathbf{r}) = \begin{bmatrix} \psi_{\mathbf{k}}^1(\mathbf{r}) & \psi_{\mathbf{k}}^2(\mathbf{r}) & \cdots & \psi_{\mathbf{k}}^{\ell_{N_{ac}}}(\mathbf{r}) \end{bmatrix} \begin{bmatrix} a_{\mathbf{k}}^1 \\ a_{\mathbf{k}}^2 \\ \vdots \\ a_{\mathbf{k}}^{\ell_{N_{ac}}} \end{bmatrix} \qquad (21)$$

which also requires the total sublattice row wavefunction vector be defined as

$$\psi_{\mathbf{k}}^{sub}(\mathbf{r}) = \begin{bmatrix} \psi_{\mathbf{k}}^1(\mathbf{r}) & \psi_{\mathbf{k}}^2(\mathbf{r}) & \cdots & \psi_{\mathbf{k}}^{\ell_{N_{ac}}}(\mathbf{r}) \end{bmatrix} \qquad (22)$$

while the sublattice coefficient column matrix of (17) generalize to

$$\mathbf{S}_{sub} = \begin{bmatrix} a_{\mathbf{k}}^1 \\ a_{\mathbf{k}}^2 \\ \vdots \\ a_{\mathbf{k}}^{\ell_{N_{ac}}} \end{bmatrix} \qquad (23)$$

Matrix equivalent of (21) is

$$\psi_{\mathbf{k}}(\mathbf{r}) = \psi_{\mathbf{k}}^{sub}(\mathbf{r}) \mathbf{S}_{sub} \qquad (24)$$

Continuing the generalization, the Hamiltonian tested by any number of atoms in the unit cell, by atom total wavefunctions, $\mathcal{H}_\mathbf{k}(\mathbf{r})$, and the same and mixed products of any number of atoms in the unit cell, by atom total wavefunctions, $\mathcal{S}_\mathbf{k}(\mathbf{r})$ (self matrix of the sublattice wavefunctions), are now

$$\mathcal{H}_\mathbf{k}(\mathbf{r}) = \begin{bmatrix} \psi_\mathbf{k}^{1*}(\mathbf{r}) \\ \psi_\mathbf{k}^{2*}(\mathbf{r}) \\ \vdots \\ \psi_\mathbf{k}^{N_{ac}*}(\mathbf{r}) \end{bmatrix} H \begin{bmatrix} \psi_\mathbf{k}^1(\mathbf{r}) & \psi_\mathbf{k}^2(\mathbf{r}) & \cdots & \psi_\mathbf{k}^{N_{ac}}(\mathbf{r}) \end{bmatrix} = \begin{bmatrix} \psi_\mathbf{k}^{1*}(\mathbf{r})H(\mathbf{r})\psi_\mathbf{k}^1(\mathbf{r}) & \cdots & \psi_\mathbf{k}^{1*}(\mathbf{r})H(\mathbf{r})\psi_\mathbf{k}^{N_{ac}}(\mathbf{r}) \\ \vdots & \ddots & \vdots \\ \psi_\mathbf{k}^{N_{ac}*}(\mathbf{r})H(\mathbf{r})\psi_\mathbf{k}^1(\mathbf{r}) & \cdots & \psi_\mathbf{k}^{l_{N_{ac}}*}(\mathbf{r})H(\mathbf{r})\psi_\mathbf{k}^{N_{ac}}(\mathbf{r}) \end{bmatrix} \quad (25a)$$

$$\mathcal{S}_\mathbf{k}(\mathbf{r}) = \begin{bmatrix} \psi_\mathbf{k}^{1*}(\mathbf{r}) \\ \psi_\mathbf{k}^{2*}(\mathbf{r}) \\ \vdots \\ \psi_\mathbf{k}^{N_{ac}*}(\mathbf{r}) \end{bmatrix} \begin{bmatrix} \psi_\mathbf{k}^1(\mathbf{r}) & \psi_\mathbf{k}^2(\mathbf{r}) & \cdots & \psi_\mathbf{k}^{N_{ac}}(\mathbf{r}) \end{bmatrix} = \begin{bmatrix} \psi_\mathbf{k}^{1*}(\mathbf{r})\psi_\mathbf{k}^1(\mathbf{r}) & \cdots & \psi_\mathbf{k}^{1*}(\mathbf{r})\psi_\mathbf{k}^{N_{ac}}(\mathbf{r}) \\ \vdots & \ddots & \vdots \\ \psi_\mathbf{k}^{N_{ac}*}(\mathbf{r})\psi_\mathbf{k}^1(\mathbf{r}) & \cdots & \psi_\mathbf{k}^{l_{N_{ac}}*}(\mathbf{r})\psi_\mathbf{k}^{N_{ac}}(\mathbf{r}) \end{bmatrix} \quad (25b)$$

Inserting (25) into (12) yields the integrated matrix versions,

$$\bar{\mathcal{H}}_\mathbf{k} = \iint d^2\mathbf{r}\, \mathcal{H}_\mathbf{k}(\mathbf{r}) = \iint d^2\mathbf{r} \begin{bmatrix} \psi_\mathbf{k}^{1*}(\mathbf{r}) \\ \psi_\mathbf{k}^{2*}(\mathbf{r}) \\ \vdots \\ \psi_\mathbf{k}^{N_{ac}*}(\mathbf{r}) \end{bmatrix} H(\mathbf{r}) \begin{bmatrix} \psi_\mathbf{k}^1(\mathbf{r}) & \psi_\mathbf{k}^2(\mathbf{r}) & \cdots & \psi_\mathbf{k}^{N_{ac}}(\mathbf{r}) \end{bmatrix}$$

$$= \begin{bmatrix} \iint d^2\mathbf{r}\, \psi_\mathbf{k}^{1*}(\mathbf{r})H(\mathbf{r})\psi_\mathbf{k}^1(\mathbf{r}) & \cdots & \iint d^2\mathbf{r}\, \psi_\mathbf{k}^{1*}(\mathbf{r})H(\mathbf{r})\psi_\mathbf{k}^{N_{ac}}(\mathbf{r}) \\ \vdots & \ddots & \vdots \\ \iint d^2\mathbf{r}\, \psi_\mathbf{k}^{N_{ac}*}(\mathbf{r})H(\mathbf{r})\psi_\mathbf{k}^1(\mathbf{r}) & \cdots & \iint d^2\mathbf{r}\, \psi_\mathbf{k}^{l_{N_{ac}}*}(\mathbf{r})H(\mathbf{r})\psi_\mathbf{k}^{N_{ac}}(\mathbf{r}) \end{bmatrix} \quad (26a)$$

$$\bar{\mathcal{S}}_\mathbf{k} = \iint d^2\mathbf{r}\, \mathcal{S}_\mathbf{k}(\mathbf{r}) = \iint d^2\mathbf{r} \begin{bmatrix} \psi_\mathbf{k}^{1*}(\mathbf{r}) \\ \psi_\mathbf{k}^{2*}(\mathbf{r}) \\ \vdots \\ \psi_\mathbf{k}^{N_{ac}*}(\mathbf{r}) \end{bmatrix} \begin{bmatrix} \psi_\mathbf{k}^1(\mathbf{r}) & \psi_\mathbf{k}^2(\mathbf{r}) & \cdots & \psi_\mathbf{k}^{N_{ac}}(\mathbf{r}) \end{bmatrix}$$

$$= \begin{bmatrix} \iint d^2\mathbf{r}\, \psi_\mathbf{k}^{1*}(\mathbf{r})\psi_\mathbf{k}^1(\mathbf{r}) & \cdots & \iint d^2\mathbf{r}\, \psi_\mathbf{k}^{1*}(\mathbf{r})\psi_\mathbf{k}^{N_{ac}}(\mathbf{r}) \\ \vdots & \ddots & \vdots \\ \iint d^2\mathbf{r}\, \psi_\mathbf{k}^{N_{ac}*}(\mathbf{r})\psi_\mathbf{k}^1(\mathbf{r}) & \cdots & \iint d^2\mathbf{r}\, \psi_\mathbf{k}^{l_{N_{ac}}*}(\mathbf{r})\psi_\mathbf{k}^{N_{ac}}(\mathbf{r}) \end{bmatrix} \quad (26b)$$

The *ij* matrix element of $\mathcal{H}_\mathbf{k}(\mathbf{r})$, $\mathcal{H}_\mathbf{k}^{ij}(\mathbf{r})$, is obtained from (25a) by inspection,

$$\mathcal{H}_\mathbf{k}^{ij}(\mathbf{r}) = \psi_\mathbf{k}^{i*}(\mathbf{r})H(\mathbf{r})\psi_\mathbf{k}^j(\mathbf{r}) \quad (27a)$$

Similarly by inspection, the *ij* matrix element of $\mathcal{S}_\mathbf{k}(\mathbf{r})$, $\mathcal{S}_\mathbf{k}^{ij}(\mathbf{r})$, is obtained from (25b),

$$\mathcal{S}_\mathbf{k}^{ij}(\mathbf{r}) = \psi_\mathbf{k}^{i*}(\mathbf{r})\psi_\mathbf{k}^j(\mathbf{r}) \quad (27b)$$

From (12), taking the $ij^{th}$ matrix element under the double spatial integration

$$\bar{\mathcal{H}}_\mathbf{k}^{ij} = \iint d^2\mathbf{r}\, \mathcal{H}_\mathbf{k}^{ij}(\mathbf{r}) \quad ; \quad \bar{\mathcal{S}}_\mathbf{k}^{ij} = \iint d^2\mathbf{r}\, \mathcal{S}_\mathbf{k}^{ij}(\mathbf{r}) \quad (28)$$

### Eigenenergies, Fermi Velocities, Overlap & Hopping Integrals

Now we know from special relativity [4] that for the low mass case compared to a high momentum $p$,

$$E = \pm \lim_{m \to small} \sqrt{c^2 p^2 + m^2 c^4} = \pm cp \lim_{m \to small} \sqrt{1 + m^2 c^2 / p^2}$$
$$= \pm cp \{1 + m^2 c^2 / (2p^2)\} \quad (29)$$

In the limit of vanishing mass $m$,

$$\lim_{m \to small} E = \pm \lim_{m \to 0} \sqrt{c^2 p^2 + m^2 c^4} = \pm cp \lim_{m \to 0} \sqrt{1 + m^2 c^2 / p^2} \approx \pm cp \quad (30)$$

Note the form of this last relationship, showing that the energy-momentum ratio is

$$\frac{\left|\lim_{m \to small} E\right|}{p} \approx c \quad (31)$$

Now by an analytical tight-binding approach, outlined in the previous section, one can write the energy about a K-point in graphene as

$$\varepsilon_q^\lambda = v_F^\varepsilon \hbar \lambda q \quad (32)$$

where $\lambda$ ($\lambda = \pm 1$) determines whether one is in the upper or lower Dirac cone, $\hbar$ is Planck's constant, $q$ the momentum magnitude about the K-point, and $v_F^\varepsilon$ the energy determined Fermi velocity. Here the energy-momentum ratio is

$$\frac{\left|\varepsilon_q^\lambda\right|}{\hbar q} = v_F^\varepsilon \quad (33)$$

obtained by looking at momentum values close to the K-point.

In order to more directly compare the relativistic (31) and non-relativistic Schrodinger equation based result (33), one can rewrite (32) using momentum as Planck's constant times $\mathbf{q}$,

$$\varepsilon_p^\lambda = v_F^\varepsilon \lambda p_q \quad (34)$$

Here energy derived Fermi velocity is given as

$$v_F^\varepsilon = -\frac{3t a_{CC}}{2\hbar} \quad (35)$$

and dependent on the nearest neighbor hopping integral $t$. Here $a_{CC}$ is the carbon-carbon closet atomic distance. For $t < 0$, $v_F^\varepsilon$ will be positive, and the band index, with values $\lambda = \pm 1$, will select the upper Dirac cone for $\lambda = +1$, and the lower Dirac cone for $\lambda = -1$. For materials with $t > 0$, the minus sign would be moved out to the $\lambda$ factor, giving - 1, and it would be relabeled as $\lambda' = -\lambda$, so that the old $-1$ would give a positive $\lambda' = -\lambda = -(-1) = +1$. The result of this process is to generate two energies, the positive "+" and the negative "-", as stated herein

$$\varepsilon_+^q = + v_F^\varepsilon p_\mathbf{q} \quad ; \quad \varepsilon_-^q = - v_F^\varepsilon p_\mathbf{q} \tag{36}$$

One immediately sees that the relativistic energies $E_+ = cp$ and $E_- = -cp$ are exactly analogous to the graphene upper and lower Dirac cone energies $\varepsilon_+^q$ and $\varepsilon_-^q$.

The nearest neighbor hopping integral $t$ is given by the expression

$$t = t_\mathbf{k}^{AB_3} = \Delta V_{sub}^a \iint d^2\mathbf{r}\, \phi_a^{A*}(\mathbf{r})\phi_a^B(\mathbf{r} + \mathbf{r}_{AB_3}) \tag{37}$$

where $\phi_a^A(\mathbf{r})$ and $\phi_a^B(\mathbf{r})$ are the atomic wavefunctions of the sublattice atoms of the types A and B. $\Delta V_{sub}^a$ is an averaged characteristic potential energy seen by the electron in the sublattice system.

It is noted here that another type of Fermi velocity can be determined, and is given by

$$v_F^H = - \frac{3(\bar{t} + \bar{s}\varepsilon_0)a_{CC}}{2\hbar} \tag{38}$$

which utilizes both a normalized hopping integral $\bar{t}$ (replace $t$ by $\bar{t}$) and a normalized self or overlap integral $\bar{s}$ (replace $s$ by $\bar{s}$), given by

$$\bar{t} = \Delta V_{sub}^a \iint d^2\mathbf{r}\, \bar{\phi}_a^{A*}(\mathbf{r})\bar{\phi}_a^B(\mathbf{r} + \mathbf{r}_{AB_3}) = t\zeta^2 \tag{39a}$$

$$\bar{s} = \iint d^2\mathbf{r}\, \bar{\phi}_a^{A*}(\mathbf{r})\bar{\phi}_a^B(\mathbf{r} + \mathbf{r}_{AB_3}) = s\zeta^2 \tag{39b}$$

$$s_\mathbf{k}^{AA} = \left(|\gamma_\mathbf{k}|^2 - 3\right)\iint d^2\mathbf{r}\, \phi_a^{A*}(\mathbf{r})\phi_a^A(\mathbf{r} - \mathbf{a}_j) = \zeta^{-2} \tag{40}$$

$$|\gamma_\mathbf{k}|^2 = \gamma_\mathbf{k}\gamma_\mathbf{k}^*$$

$$= \left(1 + e^{i\mathbf{k}\cdot\mathbf{a}_2} + e^{i\mathbf{k}\cdot\mathbf{a}_3}\right)\left(1 + e^{i\mathbf{k}\cdot\mathbf{a}_2} + e^{i\mathbf{k}\cdot\mathbf{a}_3}\right)^* \tag{41}$$

$$= 3 + 2\sum_{j=1}^{3}\cos(\mathbf{k}\cdot\mathbf{a}_j)$$

Here $\mathbf{a}_j$ are direct space lattice shift vectors, with $j = 1, 2, 3$.

Energy based Fermi velocity value for graphene in the tight binding approximation, is using (35), working in the cgs system of units,

$$v_F^\varepsilon = 9.707\times 10^7 \; cm/sec \tag{42}$$

Because both the energy and Hamiltonian based Fermi velocities have the same format [see (35) and (38)], the energy based one was used to estimate a value since it is the simplest. Equation (42) is based upon the following values of constants and unit conversions: $h = 6.626\times 10^{-27}\; erg\cdot sec$, $\hbar = h/(2\pi)$, $a_{CC} = 1.42\; Å$, $1\; Å = 10^{-8}\; cm$, $1\; eV = 1.602\times 10^{12}\; erg$, $t \approx -3\; eV$. Rounding off digits, the Fermi velocity in graphene is roughly

$$v_F^\varepsilon \approx 10^8\; cm/sec \tag{43}$$

To put this tight binding Fermi velocity value for graphene in context, refer back to earlier work on electron transport in GaAs material [5] – [8], which give saturated drift and peak velocities at room temperature of

$$GaAs: \; v_{sat\;drift} \approx 0.85\times 10^7 \; cm/sec \;\; ; \;\; v_{peak}(E = 3.2\; kV/cm) \approx 2.2\times 10^7 \; cm/sec \tag{44a}$$

For InAs material [5], [9],

$$InP: \; v_{sat\;drift} \approx 0.95\times 10^7 \; cm/sec \;\; ; \;\; v_{peak}(E = 10.5\; kV/cm) \approx 2.5\times 10^7 \; cm/sec \tag{44b}$$

For silicon, again at room temperature, the saturated drift velocity is about
$$Si: \quad v_{sat\ drift} \approx 10^7\ cm/sec \tag{45}$$
So the graphene Fermi velocity (15) is about an order of magnitude larger as seen in either silicon (17), GaAs or InP (16). Although it is a large value, and one of the main reasons for using graphene material in solid state devices today, it is nevertheless quite a finite value. To see this, just compare the Fermi velocity in graphene to that of the velocity of light in vacuum.
$$\frac{v_F^\varepsilon}{c} \approx \frac{10^8\ cm/sec}{3\times 10^{10}\ cm/sec} \approx 3\times 10^{-3} \tag{46}$$
This ratio shows that the electron velocity in graphene is both small and not relativistic.

There are other Fermi velocities which can be found. For example, again looking at momentums about the Dirac point, the Hamiltonian can be expressed as
$$\mathcal{H}_q = -v_{Fnnn}\left(\frac{3a_{CC}\hbar}{2}\right)q^2 I + v_F^H \frac{a_{CC}\hbar}{4}\left[\xi 6 q_x q_y \sigma_y - \left(q_x^2 + 3q_y^2\right)\sigma_x\right] \tag{47}$$
where $\sigma_x$ and $\sigma_y$ are Pauli spin matrices. Here the $v_{Fnnn}$ Fermi velocity is
$$v_{Fnnn} = -\frac{3 t_{nnn}^{AA} a_{CC}}{2\hbar} \tag{48}$$
which is based upon the next nearest neighbor hopping integral $t_{nnn}^{AA}$. Because $t_{nnn}^{AA}$ is about an order of magnitude smaller than the nearest neighbor hopping integral $t$ in (35), its associated Fermi velocity $v_{Fnnn}$ will also be so reduced. The true Fermi velocity to examine is $v_F^H$, given above in (38), which hits the second term in the Hamiltonian in (47).

Eigenenergy associated with Hamiltonian (47) is
$$\varepsilon_q^\lambda = -\frac{3}{2} v_F^{nnn} \hbar a_{CC} q^2 + v_F^\varepsilon \hbar \lambda q \left[1 - \xi \frac{a_{CC} q}{4} \cos 3\varphi_\mathbf{q}\right] \tag{49}$$
Notice that (49) has been written in a form where the momentum components $q_x$ qnd $q_y$ have been converted to angular representation. The trigonal warping effect is then seen explicitly, and results in a second order correction in $q$ to the eigenenergy. However, since eigenenergy to first order in $q$ is proportional to $q$, the actual correction is just the second term in square brackets, namely, $a_{CC} q \cos(3\varphi_\mathbf{q})/4$.

**Eigenvectors Based Upon 2-Spinors**

One wonders, since the eigenenergy has been examined, what one obtains for the eigenvectors. At the outset, it must be stated that the nature of the relativistic solution to the Dirac equation requires construction of 4-spinors [10], and therein lies the richness of its physics. In contrast, the solution to the non-relativistic Schrodinger equation allows for use of 2-spinors in hexagonal planar 2D materials. However, it is possible to reduce the Dirac equation in an extreme limit to a 2-spinor form. That will be done here, to highlight the original thought process, and how it should be reassessed in view of today's search for more innovative materials. In the explicit constant form [11], the Dirac equation is written as

$$\left(ic\hbar\gamma^{\mu}\partial_{\mu} - mc^2\right)\psi(x) = 0 \quad ; \quad \psi(x) = \begin{bmatrix} \psi_L(x) \\ \psi_R(x) \end{bmatrix} \tag{50}$$

(repeated indices indicate summation) with partial derivative operator matrix coefficients in chiral or Weyl representation given by

$$\gamma^0 = \begin{pmatrix} 0 & I_{2\times 2} \\ I_{2\times 2} & 0 \end{pmatrix} \quad ; \quad \gamma^i = \begin{pmatrix} 0 & \sigma^i \\ -\sigma^i & 0 \end{pmatrix} \tag{51}$$

and the $\sigma^j$ the Pauli matrices

$$\sigma^1 = \begin{pmatrix} 0 & 1 \\ 1 & 0 \end{pmatrix} \quad ; \quad \sigma^2 = \begin{pmatrix} 0 & -i \\ i & 0 \end{pmatrix} \quad ; \quad \sigma^3 = \begin{pmatrix} 1 & 0 \\ 0 & -1 \end{pmatrix} \tag{52}$$

Decomposition of $\psi$ into the left-handed $\psi_L$ and right-handed $\psi_R$ Weyl 2-spinors is consistent with the 4-spinor form of (50). They have certain rotation and boost properties which will not be delved into here. Derivative 4-operator $\partial_\mu$ has the following definitions in time and space,

$$\partial_0 = \frac{\partial}{\partial x^0} = \frac{\partial}{\partial(ct)} \quad ; \quad \partial_i = \nabla_i \quad \text{with} \quad \partial_i = \frac{\partial}{\partial x^i} \tag{53}$$

From (50) – (52), utilizing the Weyl 2-spinors, and explicitly writing out the terms for closer inspection,

$$\left(ic\hbar\left\{\begin{bmatrix} 0 & \partial_0 \\ \partial_0 & 0 \end{bmatrix} + \begin{bmatrix} 0 & \sigma^i\nabla_i \\ -\sigma^i\nabla_i & 0 \end{bmatrix}\right\} - mc^2\begin{bmatrix} I_{2\times 2} & 0 \\ 0 & I_{2\times 2} \end{bmatrix}\right)\begin{pmatrix} \psi_L(x) \\ \psi_R(x) \end{pmatrix} = 0 \tag{54}$$

and noting that $\sigma^i \nabla_i = \vec{\sigma}\cdot\vec{\nabla}$, (54) becomes,

$$\begin{bmatrix} -mc^2 & ic\hbar\left(\partial_0 + \vec{\sigma}\cdot\vec{\nabla}\right) \\ ic\hbar\left(\partial_0 - \vec{\sigma}\cdot\vec{\nabla}\right) & -mc^2 \end{bmatrix}\begin{pmatrix} \psi_L(x) \\ \psi_R(x) \end{pmatrix} = 0 \tag{55}$$

Equation (55) mixes the Lorentz group representations $\psi_L$ and $\psi_R$. For completely massless particles, the diagonal matrix elements go perfectly to zero, and a decoupling of the two implicitly stored equations in (55) occurs. This might is acceptable for a perfectly massless particle such as a photon or graviton, or reasonable for a nearly massless particle such a neutrino of a particular flavor, now known to possess an extremely tiny but finite mass. However, for an electron, with a huge mass compared to a neutrino, this becomes a questionable assumption. However, in view of the nearly 2000:1 ratio between the proton and electron masses, and to obtain 2-spinor versions of the Dirac equation directly relatable to the non-relativistic Schrodinger equation, proceed on in this direction. Coupled equations in (55) are

$$-mc^2\psi_L(x) + ic\hbar\left(\partial_0 + \vec{\sigma}\cdot\vec{\nabla}\right)\psi_R(x) = 0 \tag{56a}$$

$$ic\hbar\left(\partial_0 - \vec{\sigma}\cdot\vec{\nabla}\right)\psi_L(x) - mc^2\psi_R(x) = 0 \tag{56b}$$

Decoupling occurs with $m \equiv 0$. Then (56) become,

$$c\hbar\left(\partial_0 + \vec{\sigma}\cdot\vec{\nabla}\right)\psi_R(x) = 0 \tag{57a}$$

$$c\hbar\left(\partial_0 - \vec{\sigma}\cdot\vec{\nabla}\right)\psi_L(x) = 0 \tag{57b}$$

Since either equation is in 2-spinor form, one should be able to use either to form an analogy with graphene. For convenience, choose the $\psi_R(x)$ equation, and using (53) to write it with explicit time derivative dependence:

$$\hbar\frac{\partial \psi_R(x)}{\partial t} + c\hbar\vec{\sigma}\cdot\vec{\nabla}\psi_R(x) = 0 \tag{58}$$

For plane wave dependence of $\psi_R(x)$,

$$\psi_R(x) = \psi_{R0}e^{i(\vec{k}\cdot\vec{x}-E_R t/\hbar)} \tag{59}$$

Placing (59) into (58),

$$\hbar\left(-i\frac{E_R}{\hbar}\right)\psi_R(x) + c\hbar\vec{\sigma}\cdot i\vec{k}\psi_R(x) = 0 \rightarrow H_{R\vec{k}}\psi_{R0} = E_R\psi_{R0} \; ; \; H_{R\vec{k}} = c\hbar\vec{\sigma}\cdot\vec{k} \tag{60}$$

Momentum $\vec{q}$ about the two distinct and unique Dirac points (all others arise from these two $\vec{K}_r = \left(4\pi/\left[3\sqrt{3}\right]\right)\hat{x}$ and $\vec{K}'_l = -\vec{K}_r$), relates to Dirac points by

$$\vec{k} = \pm\vec{K}_r + \vec{q} \tag{61}$$

Inserting (61) into (60) yields

$$H_{R\vec{k}} = c\hbar\vec{\sigma}\cdot(\pm\vec{K}_r + \vec{q}) = \pm c\hbar\vec{\sigma}\cdot\vec{K}_r + c\hbar\vec{\sigma}\cdot\vec{q} \tag{62a}$$

so that the Hamiltonian about either Dirac point is given by the shifted value

$$H_{R\vec{k}}^{sh\pm} = H_{R\vec{k}} \mp c\hbar\vec{\sigma}\cdot\vec{K}_r = c\hbar\vec{\sigma}\cdot\vec{q} \tag{62b}$$

which noting that graphene viewed as strictly 2D planar has $k_z = 0$, making the Pauli matrix – vector product in (62b),

$$\vec{\sigma}\cdot\vec{q} = \begin{pmatrix} 0 & [q_x - iq_y] \\ [q_x + iq_y] & 0 \end{pmatrix} = q\begin{pmatrix} 0 & e^{-i\theta_q} \\ e^{i\theta_q} & 0 \end{pmatrix} \tag{63}$$

with $\tan(\theta_q) = q_x/q_y$ or $\theta_q = \arctan(q_x/q_y)$.

Shifted form of the governing equation for $\psi_R$ is using (60), (62) and (63),

$$H_{R\vec{q}}^{sh\pm}\psi_{R0} = E_R^{sh}\psi_{R0} \rightarrow \begin{bmatrix} -E_R^{sh} & c\hbar q e^{-i\theta_q} \\ -c\hbar q e^{i\theta_q} & -E_R^{sh} \end{bmatrix}\begin{pmatrix} \psi_{R0u} \\ \psi_{R0l} \end{pmatrix} = 0 \tag{64}$$

The determinant of this equation must be zero for a solution to exist, giving the eigenenergy solution

$$\det\begin{bmatrix} -E_R^{sh} & c\hbar e^{-i\theta_q} \\ -c\hbar e^{i\theta_q} & -E_R^{sh} \end{bmatrix} = 0 \Rightarrow \left(E_R^{sh}\right)^2 = (c\hbar)^2 q^2 \Rightarrow E_R^{sh} = \pm c\hbar q \tag{65}$$

a result found in (30). Letting $c \rightarrow v_F$ reproduces in (65) the linear energy vs momentum ($p = \hbar q$) graphene first order result for the upper $\pi^*$ (energy > 0) and lower $\pi$ (energy <

0) bands [see (34)]. For the eigenvector, write out the two spinor component requirements from (64),

$$-E_R^{sh}\psi_{R0u}+c\hbar q e^{-i\theta_q}\psi_{R0l} = 0 \Rightarrow \psi_{R0l} = \frac{E_R^{sh}}{c\hbar q}e^{i\theta_q}\psi_{R0u} \quad (66a)$$

$$c\hbar q e^{i\theta_q}\psi_{R0u} - E_R^{sh}\psi_{R0l} = 0 \quad (66b)$$

Using the first of this pair of equations (as they must contain the same information), and (65),

$$\psi_{Rq0}=\begin{pmatrix}\psi_{Rq0u}\\ \psi_{Rq0l}\end{pmatrix}=\begin{pmatrix}\psi_{Rq0u}\\ \pm e^{i\theta_q}\psi_{Rq0u}\end{pmatrix}=\begin{pmatrix}1\\ \pm e^{i\theta_q}\end{pmatrix}\psi_{Rq0u}=\frac{1}{\sqrt{2}}\begin{pmatrix}1\\ \pm e^{i\theta_q}\end{pmatrix} \quad (67)$$

Last form is for the normalized case, with ± signs associated with the upper/lower bands of graphene. If one factors out $e^{i\theta_q/2}$, rescales, and renormalizes, one obtains,

$$\psi_{Rq0}=\begin{pmatrix}1\\ \pm e^{i\theta_q}\end{pmatrix}\psi_{Rq0u}=\begin{pmatrix}e^{i\theta_q/2}\\ \pm e^{i\theta_q/2}\end{pmatrix}[e^{i\theta_q/2}\psi_{Rq0u}]=\begin{pmatrix}e^{-i\theta_q/2}\\ \pm e^{i\theta_q/2}\end{pmatrix}\overline{\psi}_{Rq0u}=\frac{1}{\sqrt{2}}\begin{pmatrix}e^{-i\theta_q/2}\\ \pm e^{i\theta_q/2}\end{pmatrix} \quad (68)$$

The last form is seen to be that of (20) of Section II. Elementary Electronic Properties of Graphene, B. Dirac Fermions, for the **K** point; and further it is noted that under Chiral Tunneling and Klein paradox, our form (67) appears as their (24) from a gauge transformation [12]; see also [13].

The $\vec{K}'_l$ point solution is easily found from the $\mathbf{K}_r$ point solution as follows. Examine the full $\vec{\sigma}\cdot\vec{k}$ product appearing in (60), enlist (61), and inspect,

$$\vec{\sigma}\cdot\vec{k} = \begin{pmatrix}0 & [k_x-ik_y]\\ [k_x+ik_y] & 0\end{pmatrix} = \begin{pmatrix}0 & \pm K_r+qe^{-i\theta_q}\\ \pm K_r+qe^{i\theta_q} & 0\end{pmatrix} \quad (69)$$

which gives the two $\vec{\sigma}\cdot\vec{k}$ products

$$\vec{\sigma}\cdot\vec{k}\Big|_{\vec{K}_r} = \begin{pmatrix}0 & K_r+qe^{-i\theta_q}\\ K_r+qe^{i\theta_q} & 0\end{pmatrix}; \vec{\sigma}\cdot\vec{k}\Big|_{\vec{K}'_l} = \begin{pmatrix}0 & -K_r+qe^{i\theta_q}\\ -K_r+qe^{-i\theta_q} & 0\end{pmatrix} \quad (70)$$

if $\vec{K}_r \to \vec{K}'_l = -\vec{K}_r = -K_r\hat{x}$ and if $q_x \to -q_x$ ($\theta_q \to \arctan(-q_x/q_y) = -\theta_q$). Then

$$\psi_{Rq0}^{\vec{K}'_l}=\frac{1}{\sqrt{2}}\begin{pmatrix}e^{i\theta_q/2}\\ \pm e^{-i\theta_q/2}\end{pmatrix} \quad (71)$$

**Eigenvectors Based Upon 4-Spinors**

If one tries to upgrade the graphene to a 4-spinor and directly compare with the fully coupled Dirac wavefunctions $\psi_L$ and $\psi_R$, one finds that no direct association is possible for the spinors, as will be demonstrated below. This is not surprising, especially in light of what had to be assumed to craft the relativistic equation into a less rich form. These assumptions included some very drastic measures, such as dropping explicit mass terms, and collapsing the dimensionality into one less dimension to get the idealized graphene sheet in 2D.

In contrast, the solution to the non-relativistic Schrodinger equation allows for use of 2-spinors. It is possible to upgrade the 2-spinor solution for the hexagonal 2D crystal case, into a 4-spinor form, by employment of the attributes of the two types of Dirac crystallographic points in **k**-space. Details of that procedure are shown elsewhere [14], and here one merely utilizes the solutions found. First consider the upper Dirac cone for graphene, and the plane wave solution of the Dirac equation for z-directed forward wave propagation. The eigenvectors for positive energy graphene ($\mathbf{K}_r$ and $\mathbf{K}'_l$, points; normalized forms) and positive energy Dirac plane wave are (lower $l$, upper $u$, spinor parts; unnormalized forms) respectively,

$$\Psi^{Graph+}_{\mathbf{K}_r, \mathbf{q}} = \frac{1}{\sqrt{2}} \begin{bmatrix} 1 \\ -e^{i\phi_q} \\ 0 \\ 0 \end{bmatrix} \Leftrightarrow \psi^{D+, tiny\ m}_l = \begin{pmatrix} 0 \\ \sqrt{2cp|_z} \\ 0 \\ 0 \end{pmatrix} \quad (72a)$$

$$\Psi^{Graph+}_{\mathbf{K}'_l, \mathbf{q}} = \frac{1}{\sqrt{2}} \begin{bmatrix} 0 \\ 0 \\ 1 \\ e^{i\phi_q} \end{bmatrix} \Leftrightarrow \psi^{D+, tiny\ m}_u = \begin{pmatrix} 0 \\ 0 \\ \sqrt{2cp|_z} \\ 0 \end{pmatrix} \quad (72b)$$

It is seen immediately in (72a) that our $\mathbf{K}_r$ graphene spinor has two upper elements, whereas the relativistic eigenvector, or 4-spinor, has a single 2$^{nd}$ element occupied. Here $\phi_q$ is angle measured by $q_x$ and $q_y$ components. So there is not an exact matchup. For the $\mathbf{K}'_l$ point [see (72b)], graphene spinor has two lower elements, whereas the relativistic eigenvector, or 4-spinor, has a single 3$^{rd}$ element occupied. So again there is not an exact matchup.

What about the lower or negative energy Dirac cone for graphene? Again using the same **K** symmetry points for graphene as for the positive energy solution, and using a relativistic eigenvector solution employing a backward going plane wave,

$$\Psi^{Graph-}_{\mathbf{K}_r, \mathbf{q}} = \frac{1}{\sqrt{2}} \begin{bmatrix} 1 \\ e^{i\phi_q} \\ 0 \\ 0 \end{bmatrix} \Leftrightarrow \psi^{D-, tiny\ m, back}_{u,} = \begin{pmatrix} \sqrt{2cp|_z} \\ 0 \\ 0 \\ 0 \end{pmatrix} \quad (73a)$$

$$\Psi^{Graph-}_{\mathbf{K}'_l, \mathbf{q}} = \frac{1}{\sqrt{2}} \begin{bmatrix} 0 \\ 0 \\ 1 \\ -e^{i\phi_q} \end{bmatrix} \Leftrightarrow \psi^{D-, tiny\ m, back}_l = \begin{pmatrix} 0 \\ 0 \\ 0 \\ \sqrt{(2cp|_z)} \end{pmatrix} \quad (73b)$$

One sees that the mismatch in elements occupied occurs again between the non-relativistic graphene eigenvector solutions ($\mathbf{K}_r$ and $\mathbf{K'}_l$ points) and the relativistic Dirac eigenvector solution (upper and lower). That is, the graphene eigenvector has two elements, whereas the relativistic eigenvector has a single element occupied.

For forward plane wave solutions to the relativistic eigenvector, the comparisons made for non-relativistic graphene to relativistic Dirac eigenvectors are,

$$\Psi^{Graph-}_{\mathbf{K}_r, \mathbf{q}} = \frac{1}{\sqrt{2}} \begin{bmatrix} 1 \\ e^{i\varphi_q} \\ 0 \\ 0 \end{bmatrix} \Leftrightarrow \psi^{D-}_l = \begin{pmatrix} 0 \\ -\sqrt{2cp|_{z,\,new}} \\ 0 \\ 0 \end{pmatrix} \quad (74a)$$

$$\Psi^{Graph-}_{\mathbf{K'}_l, \mathbf{q}} = \frac{1}{\sqrt{2}} \begin{bmatrix} 0 \\ 0 \\ 1 \\ -e^{i\varphi_q} \end{bmatrix} \Leftrightarrow \psi^{D-}_u = \begin{pmatrix} 0 \\ 0 \\ \sqrt{2cp|_{z,\,new}} \\ 0 \end{pmatrix} \quad (74b)$$

Extending the solution of the Dirac equation to include momentum transverse to the z-direction, for positive energy case,

$$\Psi^{Graph+}_{\mathbf{K}_r, \mathbf{q}} = \frac{1}{\sqrt{2}} \begin{bmatrix} 1 \\ -e^{i\varphi_q} \\ 0 \\ 0 \end{bmatrix} \Leftrightarrow \psi^{D\perp +}_u = \begin{pmatrix} 0 \\ -|\mathbf{p}_\perp| e^{i\theta_{\mathbf{p}_\perp}}/(mc) \\ 1 \\ 0 \end{pmatrix} \quad (75a)$$

$$\Psi^{Graph+}_{\mathbf{K'}_l, \mathbf{q}} = \frac{1}{\sqrt{2}} \begin{bmatrix} 0 \\ 0 \\ 1 \\ e^{i\varphi_q} \end{bmatrix} \Leftrightarrow \psi^{D\perp +}_l = \begin{pmatrix} -|\mathbf{p}_\perp| e^{-i\theta_{\mathbf{p}_\perp}}/(mc) \\ 2p|_z /(mc) \\ 0 \\ 1 \end{pmatrix} \quad (75b)$$

It is seen that the graphene eigenvector and the relativistic upper eigenvector both have two elements. However, this is not the situation for the lower relativistic eigenvector, unless $|\mathbf{p}_\perp|/p \ll 1$ or $p_z/|\mathbf{p}_\perp| \ll 1$, making the longitudinally z-directed momentum large compared to the transverse momentum, or the reverse.

For the negative energy case,

$$\Psi^{Graph-}_{\mathbf{K}_r, \mathbf{q}} = \frac{1}{\sqrt{2}} \begin{bmatrix} 1 \\ e^{i\varphi_q} \\ 0 \\ 0 \end{bmatrix} \Leftrightarrow \psi^{D\perp -}_u = -\begin{pmatrix} 2p|_z /(mc) \\ |\mathbf{p}_\perp| e^{i\theta_{\mathbf{p}_\perp}}/(mc) \\ 1 \\ 0 \end{pmatrix} \quad (76a)$$

$$\Psi^{Graph-}_{\mathbf{K}'_l, \mathbf{q}} = \frac{1}{\sqrt{2}} \begin{bmatrix} 0 \\ 0 \\ 1 \\ -e^{i\varphi_\mathbf{q}} \end{bmatrix} \Leftrightarrow \psi_l^{D\perp-} = \begin{pmatrix} -|\mathbf{p}_\perp|e^{-i\theta_{\mathbf{p}_\perp}}/(mc) \\ 0 \\ 0 \\ 1 \end{pmatrix} \quad (76b)$$

Here the element mismatch occurs for the upper relativistic eigenvector and the graphene eigenvector. That element number mismatch goes away for either $|\mathbf{p}_\perp|/p << 1$ or $p_z/|\mathbf{p}_\perp| << 1$.

## Conclusions and Future Outlook

From the above analysis, one sees that graphene, and by analogy to it, similar 2D materials, are analogous to low mass relativistic particles in the eigenenergy sense. It is much harder to draw analogies between the eigenvectors arising from fundamentally different kinetic physics, namely, the Schrodinger nonrelativistic equation based 2D solid state materials, exemplified by monoatomic graphene, and the relativistic Dirac equation. The closest association arises for the decoupled relativistic equation and the graphene solutions. However, we know one of the main drawbacks of the simple so called condensed matter Dirac materials is that they seem to lack a band gap. A paper which thoroughly explored the behavior of graphene field effect transistors, provided hole p and electron n density profiles and dependence on Fermi energy $\varepsilon_F$, their respective drain partial current profiles $J_{D,p}$ and $J_{D,n}$ and dependence on $V_{GS}$, drain current $J_D$ and transconductance $g_m$ vs $V_{GS}$ and $V_{DS}$, gate capacitance $C_g$ vs $V_{GS}$, and the short circuit current-gain cut-off frequency $f_t$ vs $J_D$ and gate length $L$ [15]. It found that the ambipolar transport with no gap leads to limited channel pinch-off, and a lack of an off-state [15]. This does not bode well for semiconducting uses in active electronic devices such as transistors. One wonders if the linear energy vs momentum dispersion relation is retained while opening up a bandgap in some new materials? Or can one avoid completely the type of bandstructure associated with Dirac points and gapless behavior in other 2D materials? Despite these bandgap issues, there have been many uses found for graphene, too numerous to list comprehensively here, but noting two examples [16], [17].

So one wonders what are the possibilities in terms of which materials may play a role in future electronics? A 2D material related to graphene, just treated, is graphyne [18] - [21], which consists of a lattice of benzene rings connected by triple acetylene bonded C-C atoms. In comparison to graphene's pure $sp^2$ hybridization and diamond's pure $sp^3$ hybridization, graphyne can be considered a mixed hybridization, $sp^n$, where $1 < n < 2$. depending on the content of the acetylene groups. Some of its graphyne's allotropes have finite bandgaps (predicted to be between 0. 5 and 2.5 eV), useful for semiconductor electronics and getting one away from the metallic state, which is essentially what graphene is viewed from an electronic transport perspective. Graphyne also has promise in thermoelectrics, energy storage and battery usage. Another monoatomic material, consisting only of boron atoms, with some allotropes displaying bandgaps (on the order of one to a few eV) and interesting for semiconductor electronics, is borophene [22] – [26]. Some of its hexagons are filled with boron atoms, modifying its 2D hexagonal honeycomb structure.

It is informative to note the following results from classical quantum mechanics for the effective mass of materials [27],

$$\left(\frac{1}{m^*}\right)_{ij} = \frac{1}{\hbar^2}\frac{\partial^2 \varepsilon(\vec{k})}{\partial k_i \partial k_j} \tag{77}$$

Evaluate this at a Dirac point using the x-axis shifts of $\pm q_x$ about 0, (32) for the energy and $\lambda = +1$ for the upper cone. One obtains for the isotropic case

$$\frac{1}{m^*} = \frac{1}{\hbar^2}\frac{\partial^2 \varepsilon(\vec{k})}{\partial q^2} = \frac{1}{\hbar^2}\frac{(d\varepsilon/dq_x)_{q_{x1}} - (d\varepsilon/dq_x)_{-q_{x1}}}{q_{x1} - (-q_{x1})}$$
$$= \frac{1}{\hbar^2}\frac{(d[v_F^\varepsilon \hbar q_x]/dq_x)_{q_{x1}} - (d[-v_F^\varepsilon \hbar q_x]/dq_x)_{-q_{x1}}}{q_{x1} - (-q_{x1})} = \frac{v_F^\varepsilon}{\hbar q_{x1}} \tag{78}$$

which shows that in the limit $q_{x1} \to 0$ (the Dirac point momentum approaches zero), the effective mass at a Dirac point is zero, since its inverse value approaches infinity. Although completely consistent with the reduction employed in obtaining (57), this highlights the problem with 2D Dirac materials. They are very unlikely to reproduce the desirable finite bandgap properties of more conventional, but highly effective semiconductors for electronics.

Finally, although it has been suspected for some time graphene may not be perfectly thin, and this may play into its mechanical properties [28], [29], it has been definitively shown recently [30] where the in-plane phonon graphite mode frequency for both graphene and graphite follow a similar trend, expressed as
$\omega(P) = \left(1/[\pi c\sqrt{m}]\right) SQRT\{E_0\beta^2 \left(\exp[\beta r_0 a_{33} P/(c_{11}^{2D} + c_{12}^{2D})] - 1\right)\exp[\beta r_0 a_{33} P/(c_{11}^{2D} + c_{12}^{2D})]\}$. Here $r$ is the separation of the nearest C-C atoms, $r_0$ is the unstrained C-C bond length, $E_0$ and $\beta$ denote the depth and width of the potential, respectively, in a Morse potential description, and $a_{33}$ is the interlayer spacing. $a_{33}$ relates to each carbon atom having electronic orbitals (the $2p_z$ orbitals) that extend some distance above and below the graphene sheet and resist compression. Elastic constants are denoted by the usual tensor constants $c_{11}^{2D}$ and $c_{12}^{2D}$. The similarity of the trends was determined to a maximum of about 7 GPa, with shift rates of 5.4 cm$^{-1}$GPa$^{-1}$ for graphene and 4.7 cm$^{-1}$GPa$^{-1}$ for graphite, implying that monolayer graphene has similar in-plane and out-of-plane stiffnesses, and anharmonicities to graphite.

Implications for alternatives to graphene, such as graphyne, are clear, that with its mixing of the pure sp$^2$ hybridization of graphene and diamond's pure sp$^3$ hybridization, the need for 4-spinor eigenvectors may be warranted and its energy spectrum more conducive to obtaining desirable $\varepsilon(\mathbf{k})$ properties for electronics.

## Acknowledgments

This work was supported by the US Office of Naval Research.